\documentstyle[12pt]{article}
\begin{document}
\title{Scalar Field Inhomogeneous\\ Cosmologies}
\author{A. Feinstein, J. Ib\'a\~nez and P. Labraga\\
Dpto. F\'\i sica Te\'orica, Universidad del Pa\'\i s Vasco, Bilbao, Spain.}
\date{}
\maketitle
\vskip 2cm
\begin{abstract}\noindent
Some exact solutions for the Einstein field equations corresponding to
inhomogeneous $G_2$ cosmologies with an exponential-potential scalar field
which generalize solutions obtained previously are considered. Several
particular cases are studied and the properties related to generalized
inflation and asymptotic behaviour of the models are discussed.
\end{abstract}
\newpage
\newcommand{\be}{\begin{equation}}
\newcommand{\ee}{\end{equation}}
\newcommand{\case}[1]{\vskip 1cm\hskip 1cm {\em #1}\par\vskip 0.6cm}
\def\thesection       {\Roman{section}}

\section{Introduction}

It is generally believed that the scalar field with different kinds of
potentials or different couplings to the curvature makes the space-time inflate
and isotropize. In this sense inflation is thought as a sort of mechanism that
achieves the Misner's program \cite{misner}, i.e. to explain the large-scale
homogeneity of the Universe without imposing very special conditions on the
initial expansion. This belief is based on the cosmic no-hair theorem proved
by Wald \cite{wald}, which asserts that all initially expanding Bianchi
models, except type IX, with a positive cosmological constant evolve towards
the de Sitter solution. The same was later proved for a particular class of
inhomogeneous cosmological models by Jensen and Stein-Schabes \cite{JS}. It
has been pointed out \cite{barrow}, however, that these theorems exclude a
large class of cosmological models. In addition, Heusler \cite{heusler}
extended the Collins and Hawking analysis \cite{CH} to scalar field
cosmologies and showed that Bianchi models, except type IX, coupled with a
scalar field with a convex positive potential  isotropize only if they admit a
FRW model as a particular case. Actually the later result confirms that the
presence of a scalar field does not imply that the model isotropizes, and more
assumptions about either the form of the potential or the coupling between the
field and the geometry should be then explored.

Recently, there has been a considerable interest in scalar fields with the
Liouville form (exponential) potential. It arises as an effective potential in
many theories such as Jordan-Brans-Dicke theory, Salam-Sezgin theories, the
superstring theories and others \cite{halliwell}. On the other hand there exist
possibilities to find exact solutions of the Einstein equations for this
potential in a variety of cases \cite{AFI-I}, \cite{AFI-II}. Moreover, it has
been shown that the Einstein field equations for homogeneous models with
scalar field with exponential potential can be decoupled and, therefore, are
reduced to an autonomous system of first-order ordinary differential equations
which can be analysed qualitatively \cite{CHI}, with the de Sitter solution
acting as an attractor \cite{KM}, \cite{IHC}, for particular values of the
slope of the potential.

In the inhomogeneous case, little has been done on exact solutions and, due to
the dependence of the metric on the spatial coordinates, the Einstein equations
can not be reduced to an autonomous system and, therefore, the asymptotic
behaviour can not be analysed in general. Although, some numerical and
qualitative results suggest that under certain conditions, inhomogeneities can
be smoothed and spacetimes isotropize, inhomogeneous exact solutions found
recently behave in a different way \cite{AFI-II}. These models represent an
inhomogeneous generalization of Bianchi I cosmologies, where the
inhomogeneities are induced either by the irregularities of the scalar field
or by the initial inhomogeneities in the geometry, due for example to the
presence of strong gravitational waves. The behaviour of the models  suggests
that more work on inhomogeneous solutions with scalar field and, specifically,
on exact solutions is needed to clarify the dependence of the inflation and
the isotropization on the inhomogeneity of the spacetime.

It has been shown by Wainwright \cite{wainwright} that the self-similar
solutions serve as attractors in the homogeneous case, therefore playing an
important role in studying the asymptotic behaviour of the  cosmological
models.
The inhomogeneous models found in \cite{AFI-II} were self-similar. Thus, it
will
be interesting to see as to whether the models of \cite{AFI-II} serve as
attractors for the more general inhomogeneous solutions, generalizing the
property found by Wainwright into inhomogeneous cosmologies.

In this paper we generalize the results of \cite{AFI-II} finding a large class
of inhomogeneous solutions with scalar field with exponential potential. In
Section II we describe the geometry and the matter content of the model, and
we present the corresponding Einstein equations. In Section III we deal with
these equations and obtain some particular solutions, which can be classified
into two main groups. We also make a brief discussion about the existence of
inflation in those cases. We conclude and summarize our results in Section IV.

\section{The Einstein Equations}
We will consider solutions with one dimensional inhomogeneity, which
can be described by the generalized Einstein-Rosen space-times \cite{CCM}
admitting an Abelian group of isometries $G_2$ and include the
Bianchi models of type I-VII as particular cases and, therefore, the
flat and open FRW solutions. The line element is
\be
ds^2=e^f(-dt^2+dz^2)+g_{ab}dx^adx^b,\qquad a,b=1,2,
\ee
and the functions $f$ and $g_{ab}$ depend on $t$ and $z$.

Assuming for simplicity that the two Killing vectors are hypersurface
orthogonal, the line element (1) may be put in the following diagonal form
\be
ds^2=e^{f(t,z)}(-dt^2+dz^2)+G(t,z)(e^{h(t,z)}dx^2+e^{-h(t,z)}dy^2).
\ee
This line element generalizes one studied in \cite{AFI-II} by allowing the
transitivity surface area $G$ to depend on $t$ and $z$ coordinates.

The matter source of the metric is that of a minimally coupled scalar
field with the Liouville type of potential
\be
V(\phi)=\Lambda e^{k\phi}.
\ee
The energy-momentum tensor for the scalar field is
\be
T_{\alpha\beta}=\phi_{,\alpha}\phi_{,\beta}-g_{\alpha\beta}\left(
{1\over 2}\phi_{,\gamma}\phi^{,\gamma}+V(\phi)\right).
\ee
This tensor can be rewritten in the form of a perfect fluid stress-energy
tensor
\be
T_{\alpha\beta}=(\hbox{\sl p}+\rho)u_\alpha u_\beta +
\hbox{\sl p}g_{\alpha\beta},
\ee
where $u_\alpha$, $\rho$ and {\sl p} are expressed in terms of the scalar
field as follows
\be
u_\alpha={\phi_{,\alpha}\over\sqrt{-\phi_{,\gamma}\phi^{,\gamma}}},
\ee
\be
\rho=-{1\over 2}\phi_{,\alpha}\phi^{,\alpha}+V(\phi),
\ee
\be
\hbox{\sl p}=-{1\over 2}\phi_{,\alpha}\phi^{,\alpha}-V(\phi),
\ee
as long as the gradient of the scalar field is timelike. This interpretation
will be useful to study the kinematical behaviour of the solutions.

The kinematical quantities relevant to the study of these models are the
expansion
\be
\Theta=u_{\mu;\nu}\,g^{\mu\nu}
\ee
and the deceleration parameter
\be
q=-3\Theta^{-2}\left(\Theta_{;\alpha}u^\alpha+{1\over 3}\Theta^2\right).
\ee
The deceleration parameter will act as an indicator of the existence of
inflation. If $q>0$ the solution decelerates in the ``standard" way, while
$q<0$ indicates inflation.

The Einstein equations for the line element (2) and the matter
described by the stress-energy tensor (4), along with the
Klein-Gordon equation for the scalar field, are given by
\be
\ddot\phi - \phi'' + {\dot G\over G} \dot\phi - {G'\over G} \phi' +
e^{\displaystyle f} {\partial V \over \partial \phi}= 0,\label{KGE}
\ee
\be
{\ddot G\over G} - {G''\over G} = 2\,e^{\displaystyle f}\,V,
\ee
\be
\ddot h - h'' + {\dot G\over G} \dot h - {G'\over G} h' = 0,
\ee
\be
{\dot G'\over G} - {1\over 2} {\dot GG'\over G^2} + {1\over 2} \dot hh' -
{1\over 2} f'{\dot G\over G} - {1\over 2}\dot f{G'\over G}=-\dot\phi\phi',
\ee
\be
{\ddot G\over G}+{G''\over G}-{1\over 2}\left({\dot G\over G}\right)^2
-{1\over 2}\left({G'\over G}\right)^2-\dot f{\dot G\over G}-f'{G'\over G}
+{1\over 2}\dot h^2+{1\over 2}h'^2=-\dot\phi^2-\phi'^2,
\ee
where Eq.(\ref{KGE}) is the Klein-Gordon equation, and primes and dots
represent the derivatives with respect to $z$ and $t$, respectively.

We must note that Eq.(12) has the form of an inhomogeneous wave equation. On
the other hand, the equation for the transversal (or wavelike) part of the
metric, $h$, Eq.(13), is very similar to Eq.(11) for the scalar field. This
will permit us to solve the Klein-Gordon equation in terms of the metric
function $h$. Finally, Eqs.(14) and (15) are the equations for the longitudinal
part of the metric function $f$.

\section{Solving the Equations}
Without any loss of generality, we can write the scalar field in the form:
\be
\phi=-{k\over2}\log(G) + \Phi
\ee
and substitute this expression into Eq.(11). Then, using Eq.(12) and the
expression (3) for the potential, the following equation for the function
$\Phi$ results:
\be
\ddot\Phi - \Phi''+{\dot G\over G}\dot\Phi - {G'\over G}\Phi'=0.
\ee
This is the same equation as for the metric function $h$, Eq.(13), and allows
us to write the following expression for the scalar field:
\be
\phi=-{k\over 2}\log(G)+mh,\quad\quad m=\hbox{constant}.
\ee

It is convenient now to see Eq.(12) not as a differential equation for $G$,
but rather as an equation which defines the function $f$:
\be
f(t,z)=-k\,\phi+\log\left({\ddot G\over G}-{G''\over G}\right)-\log(2\Lambda).
\ee

We have used by now Eqs.(11) and (12) and are left with (13)-(15) to
determine the functions $G(t,z)$ and $h(t,z)$.

We assume now that the metric function $G$ is separable, as follows:
\be
G(t,z)=T(t)Z(z).
\ee
Due to the symmetry of the Eqs.(13)-(15) under $t$-$z$ interchange, we assume
two different behaviours for function $h$:

\begin{list}{}{\itemindent 2cm}
\item[i)] \be e^h(t,z)=Q(t)Z(z)^n\ee
\item[ii)] \be e^h(t,z)=T(t)^nP(z)\ee
\end{list}

The second ansatz is somewhat similar to that one used by Ruiz and Senovilla in
their study of inhomogeneous perfect-fluid $G_2$ cosmologies, \cite{RS}, while
the first one is allowed by the peculiar behaviour of the system of
differential equations.

For the first case, Eq.(13) takes this form:
\be
n{Z''\over Z}={\dot T\dot Q\over TQ}+{\ddot Q\over Q}-{\dot Q^{^2}\over
Q^{^2}}=n\varepsilon a^2,\qquad (\varepsilon =0,\pm 1).
\ee
while for the second case,
\be
n{\ddot T\over T}={Z'P'\over ZP}+{P''\over P}-{P'^{^2}\over P^{^2}}=
n\varepsilon a^2,\qquad (\varepsilon =0,\pm 1),
\ee

We will study both types of solutions in the following sections.

\subsection{Case i) \hskip 1cm $e^h(t,z)=Q(t)Z(z)^n$}
We are now able to write down the form of the desired solutions, in terms of
$T(t)$, $Q(t)$ and $Z(z)$, which is
\be
\left\{
\begin{array}{rcl}
G(t,z)&=&T\,Z, \\*[3pt]
h(t,z)&=&\log(Q)+n\log(Z), \\*[3pt]
\phi (t,z)&=&-{k\over 2}\log(T\,Z)+m\,h, \\*[3pt]
f(t,z)&=&-k\,\phi+\log\left({\ddot T\over T}-{Z''\over Z}\right)-
\log(2\Lambda), \\*[3pt]
V(\phi)&=&\Lambda\,e^{\displaystyle k\phi}.
\end{array}
\right.
\ee

The spatial part of Eq.(23) results in the following three different
types of solutions:
\be
Z(z)=\left\{
\begin{array}{ll}
A\cosh(az)+B\sinh(az) &\hbox{if}\quad \varepsilon=1,\\
Az+B &\hbox{if}\quad \varepsilon=0,\\
A\cos(az)+B\sin(az) &\hbox{if}\quad \varepsilon=-1.
\end{array}
\right.
\ee

On the other hand, if we take the temporal part of Eq.(23), we can formally
write a relation between the two functions, $Q(t)$ and $T(t)$, as follows:
\be
\ln Q(t)=n\varepsilon a^2\int{{\displaystyle{\left(\int_o^t{T(\tau)d\tau}
\right)}\over T(t)}dt}.
\ee

This expression, however, is not particularly useful in order to find exact
solutions, because it yields two integro-differential equations. A more
practical way of handling the equations would be to differentiate the
function $f$ in Eq.(19) and then substitute $\dot f$ and $f'$ in Eqs.(14) and
(15). This will be done below.

We can now easily obtain some particular solutions just by choosing one
of the three possible cases from (26), and then finding the corresponding
solution for $T(t)$.

\subsubsection{Particular Solutions}

\case{The Linear Case}

The simplest type of solution is obtained when the spatial dependence of the
metric functions is linear. Rescaling constants $A$ and $B$ we can assume,
without loss of generality, that $Z(z)=z$. It can be shown then that the
function $T$ is given by $T(t)=\exp\left(\tilde At+\tilde B\right)$, where
$\tilde A$ and $\tilde B$ are arbitrary constants. The line element then
takes the following form:
\be
ds^2=ze^t\left[ {1\over 2\Lambda}z^{-\sqrt{2-n^2}}\left( -dt^2+dz^2\right)+
z^ndx^2+z^{-n}dy^2\right] .
\ee
In the special case when $n^2=1$ the metric represents a homogeneous
isotropic universe.

We have looked at some invariant quantities like $R^{ab}R_{ab}$,
$R^{abcd}R_{abcd}$ and have not found any singular behaviour apart from the
irregular behaviour at $z=0$. This singularity disappears in the special
case with $n^2=1$, when the metric is of FRW type.

Making use of expressions (7) and (8), we can readily calculate the
expressions for the energy density and the pressure, resulting
\be
\rho ={{\Lambda\,{z^{-3 + {\sqrt{2 - {n^2}}}}}\,
\left( -3 + {n^2} + 2\,{\sqrt{2 - {n^2}}} + 3\,{z^2} \right) }\over
{2\,{e^t}}}
\ee

and

\be
\hbox{\sl p}={{\Lambda\,{z^{-3 + {\sqrt{2 - {n^2}}}}}\,
\left( -3 + {n^2} + 2\,{\sqrt{2 - {n^2}}} - {z^2} \right) }\over
{2\,{e^t}}}.
\ee
The expressions for the expansion and the deceleration parameter are quite
lengthy, and we will not write them down explicitly here. We will discuss the
behaviour of the deceleration parameter at the end of this section.
\newpage
\case{The Hyperbolic Case}
The second choice for the function $Z$ is, in the most general form,
\be
Z(z)=A\sinh(az)+B\cosh(az).
\ee
To simplify the equations, we will take the special case $Z(z)=\sinh(az)$. The
general study of the resulting equations for $T(t)$ is more complicated than
in the previous case, and it is given in the Appendix. There, we will show
that it is impossible to find an explicit general solution for the system of
Eqs.(11)-(15), when we suppose $Z(z)=A\sinh(az)$. In this subsection we will
present one of the possible particular solutions.

One can suppose, for instance, that the solution for $T$ could be of the same
form as for $Z$, that is, $T(t)=\tilde A\sinh(bt)+\tilde B\cosh(bt)$. If we
introduce this solution into the equations, we come to the conclusion that
$\tilde A$ must be equal to $\tilde B$, so the actual solution is again
an exponential. The complete form is
\be
\left\{
\begin{array}{rcl}
G(t,z)&=&\sinh(az)\exp(bt),\\*[3pt]
h(t,z)&=&n\log(\sinh(az))+{\displaystyle{na^2\over b}}t,\\*[8pt]
\phi(t,z)&=&C_1\log(\sinh(az))+C_2t,\\*[3pt]
f(t,z)&=&-kC_1\log(\sinh(az))-kC_2t+\log\left({\displaystyle{b^2-a^2\over
2\Lambda}} \right) ,\\*[8pt]
V(\phi)&=&\Lambda\,\exp(kC_2t)\left[\sinh(az)\right]^{kC_1},
\end{array}
\right.
\ee
\[
\left(
\begin{array}{cl}
&b=a\sqrt{{k^2+2\over k^2-2}}\\*[3pt]
\hbox{with}&C_1={(k^2-2n^2+2)^{(1/2)}-k\over 2}\\*[3pt]
&C_2={a^2(k^2-2n^2+2)^{(1/2)}-kb^2\over 2b}
\end{array}
\right).
\]

The expressions for the energy density and pressure, Eqs.(7) and (8), become
more complicated for this solution:
\be
\rho =\Lambda \exp(kC_2t)\sinh^{(kC_1)}(az)\left( {C_2^2-C_1^2a^2\over b^2-
a^2}\cot^2(az)+1\right) ,
\ee
\be
\hbox{\sl p}=\Lambda \exp(kC_2t)\sinh^{(kC_1)}(az)\left( {C_2^2-C_1^2a^2\over
b^2- a^2}\cot^2(az)-1\right) .
\ee
The corresponding line element is:
\begin{eqnarray}
ds^2&=&{b^2-a^2\over 2\Lambda}e^{-kC_2t}\sinh^{-kC_1}(az)(-dt^2+
dz^2)+\nonumber\\
&&+\left( \sinh^{1+n}(az)e^{{na^2+b^2\over b}t}dx^2+\sinh^{1-n}(az)e^{{b^2-
na^2\over b}t}dy^2\right).
\end{eqnarray}

{}From the expressions (33) and (34) we deduce that there is a periodic
singular behaviour in $z$.

\case{The Trigonometric Case}
This is the last type of solutions, which is not very different from the
previous one. We can try to find again a particular solution with an
exponential behaviour in time, as it seems to be the most common among all our
solutions. In fact, there is a particular solution of this kind, namely:
\be
\left\{
\begin{array}{rcl}
G(t,z)&=&\sin(az)\exp(bt),\\
h(t,z)&=&n\log(\sin(az))-{\displaystyle{na^2\over b}}t,\\*[5pt]
\phi(t,z)&=&K_1\log(\sin(az))-K_2t,\\
f(t,z)&=&-kK_1\log(\sin(az))+kK_2t+\log\left({\displaystyle{a^2+b^2\over
2\Lambda}}\right) ,\\*[5pt]
V(\phi)&=&\Lambda\,\exp(-kK_2t)\left[\sin(az)\right]^{kK_1},
\end{array}
\right.
\ee
\[
\left(
\begin{array}{cl}
&b=a\sqrt{{k^2+2\over 2-k^2}}\\*[3pt]
\hbox{where}&K_1={(k^2-2n^2+2)^{(1/2)}-k\over 2}\\*[3pt]
&K_2={a^2(k^2-2n^2+2)^{(1/2)}+kb^2\over 2b}
\end{array}
\right).
\]
As can be easily seen, this solution is very similar to the hyperbolic one.

\subsubsection{Do The Models Inflate?}
As to the existence of inflation in these models, we have to say that
all the expressions for the corresponding deceleration parameters are quite
complicated, and therefore, little information can be obtained from their
explicit form. However, we can look at the behaviour of $q$ against one of
the variables $t$ or $z$ for some particular values of constants. It can be
shown analytically that, in all of the cases discussed above, the sign of
the deceleration parameter does not depend on time. This means that the
inflation has a spatial character. Some of the regions, depending on their
location in the universe, will inflate forever, but some will never experience
the inflation.

The most remarkable feature is that all of the models we have looked at show
a strong dependence on the spatial coordinate $z$. Moreover, given the same
kind of solution with slightly different values for the constants, we get a
quite different behaviour of $q$ as a function of $z$. In the case of
periodic solutions ($Z=\sin(az)$) the inflating and non-inflating regions
interchange periodically.

When a cosmological model deviates strongly from homogeneity it can
be extremely difficult to see whether the model inflates or not. When this
happens, however, we can use a somewhat weaker way to deduce about the
inflationary behaviour of the model, just by looking at the fulfillment
of the strong energy condition. The breaking of this condition is known
to be a necessary for inflation to take place. Therefore, we
can study the behaviour of the quantity ${\cal E}=\rho + 3\sl p$ for
different values of the variables and the constants. When this quantity
is such that ${\cal E}\leq 0$ the strong energy condition is violated, and
inflation might take place.

If we do so for the first of our models, the linear case, using
expressions (29) and (30), we can see that $\cal E$ is always negative
(again, the sign of this magnitude has no dependence on time, since this
coordinate always appears in the expressions of $\rho$ and $\sl p$ in the
exponential form). Therefore, the model can inflate. For the special case
with $n^2=1$ we have ${\cal E}=0$, which corresponds to an equation of
state of the form ${\sl p}=-\rho /3$.

In the hyperbolic case, making use of Eqs.(33) and (34), we can see that
the strong energy condition is broken only for some values of the
coordinate $z$ and depend strongly on $k$. The energy condition is always
broken for small $z$. If $k^2\rightarrow 2$ the energy condition is broken
for a wide range of values of $z$. When $k$ grows, this range decreases
strongly. Therefore, we can say that inflation is always possible near the
hypersurface $z=0$, but is less probable in other regions.

\subsection{Case ii) \hskip 1cm $e^h(t,z)=T(t)^nZ(z)$}
As we have discussed at the beginning of this section, we can use a different
assumption for the form of the metric function $h$, using Eq.(22) instead of
Eq.(21). If we do so, we must deal with Eq.(24) rather than Eq.(23):
\[
n{\ddot T\over T}={Z'P'\over ZP}+{P''\over P}-{P'^{^2}\over P^{^2}}=
n\varepsilon a^2,\qquad (\varepsilon =0,\pm 1),
\]
and again we can easily solve the temporal part of this equation, which is
equivalent to Eq.(26), but in this case with $t$ as the variable:
\be
T(t)=\left\{
\begin{array}{ll}
A\cosh(at)+B\sinh(at) &\hbox{if}\quad \varepsilon=1,\\
At+B &\hbox{if}\quad \varepsilon=0,\\
A\cos(at)+B\sin(at) &\hbox{if}\quad \varepsilon=-1.
\end{array}
\right.
\ee
Once more we have three different cases.

Similarly to the previous case, we can write down the general form of the
solutions, by taking $G$ from Eq.(20), $h$ from Eq.(22), and $\phi$, $f$
and $V$ from Eqs.(18), (19) and (3) respectively. The most general solution
can be written then in the following form:
\be
\left\{
\begin{array}{rcl}
G(t,z)&=&T\,Z, \\*[3pt]
h(t,z)&=&n\log(T)+\log(P), \\*[3pt]
\phi (t,z)&=&-{k\over 2}\log(T\,Z)+m\,h, \\*[3pt]
f(t,z)&=&-k\,\phi+\log\left({\displaystyle{\ddot T\over T}-{Z''\over Z}}
\right)-\log(2\Lambda), \\*[3pt]
V(\phi)&=&\Lambda\,e^{\displaystyle k\phi}.
\end{array}
\right.
\ee
Again, we can try to find the most general solution corresponding to
each particular case of $T(t)$, or at least, find some particular solutions
that could be of interest in each case.
\newpage
\subsubsection{Particular solutions}

\case{The Linear Case}
If we take the simplest case from Eq.(37), which is the linear solution
for $T(t)$, we obtain a general solution analogous to Eq.(28), that is,
\be
ds^2=te^{\displaystyle z}(-dt^2+dz^2+t^{\sqrt{2}}\,dx^2+
t^{-\sqrt{2}}\,dy^2).
\ee
The expressions for the energy density and the pressure are now
\be
\rho={1\over 4}e^{-z}\left( t^{-3}-3t^{-1}\right) ,
\ee

and

\be
\hbox{\sl p}={1\over 4}e^{-z}\left( t^{-3}+t^{-1}\right) .
\ee

This is an interesting solution owing to its simplicity, which allows us to
calculate simple expressions for the expansion and the deceleration parameter:
\be
\Theta={{3 {t^4} - 4 {t^2} + 3}\over
{2 {e^{\displaystyle z\over 2}} {t^6} {{\left( {t^{-3} - t^{-1}}\right) }^
{{3\over 2}}}}},
\ee
\be
q=\Theta^{-2} {{\left( 3 {t^2} -1 \right)  \left( {t^4} + 9 \right) }\over
{2 {e^{\displaystyle z}} {{\left( t + 1 \right) }^3} {t^3} {{\left( t - 1
\right) }^3}}}.
\ee
Looking at the sign of $q$, we can now easily see that inflation starts at
\be
t=\frac{1}{\sqrt{3}}.
\ee
This universe expands in the standard way before $t=1/\sqrt{3}$ and then
inflates.
\newpage
\case{The Hyperbolic Case}
If we consider the second case, $T(t)\propto\sinh(at)$, we can take a
particular solution which is the equivalent of equation (32), interchanging
variables. The spatial part of function $G(t,z)$ is an exponential, as in
the previous case. All the particular solutions found appear to have an
exponential behaviour in $z$. The solution is the following:
\be
\left\{
\begin{array}{rcl}
G(t,z)&=&\sinh(at)\exp(bz),\\
h(t,z)&=&\log(\sinh(at))+{\displaystyle{a^2\over b}}z,\\
\phi(t,z)&=&-k\log(\sinh(at))-{\displaystyle{k\over 2}}\left(
{\displaystyle{a^2+b^2\over b}} \right) z,\\
f(t,z)&=&k^2\log(\sinh(at))+{\displaystyle{k^2\over 2}}\left(
{\displaystyle{a^2+b^2\over b}} \right) z+\log\left({\displaystyle{a^2-b^2\over
2\Lambda}}\right),\\
V(\phi)&=&\Lambda\,e^{\displaystyle k\phi},
\end{array}
\right.
\ee
\[
\left(\hbox{with }b=a\sqrt{{k^2+2\over k^2-2}}\right).
\]

The expressions for the energy density and the pressure are as follows:
\be
\rho={e^{\left[-{k^2\over 2}\left({a^2+b^2\over b}\right)z\right]}
\over (\sinh(at))^{k^2}} \left[{a^2-b^2\over 4\Lambda}\left(k^2a^2\coth^2(at)
-{k^2\over 4}\left({a^2+b^2\over b}\right)^2\right) +\Lambda\right] ,
\ee
\be
\hbox{\sl p}={e^{\left[-{k^2\over 2}\left({a^2+b^2\over b}\right)z\right]}
\over (\sinh(at))^{k^2}} \left[{a^2-b^2\over 4\Lambda}\left(k^2a^2\coth^2(at)
-{k^2\over 4}\left({a^2+b^2\over b}\right)^2\right) -\Lambda\right] .
\ee

It is worthwhile to mention that among these solutions we have also found a
solution which represents a universe with a cosmological constant and a
vanishing scalar field. The corresponding line-element reads as follows
\be
ds^2={1\over\Lambda}\left( -dt^2+dz^2\right) +\sinh^2(t)dx^2+\cos^2(z)dy^2.
\ee

This line-element can be put into the following form:
$$
ds^2=\frac{1}{\Lambda}\left( -dt^2+\Lambda\sinh^2(t)dx^2\right)\, +
\,\frac{1}{\Lambda}\left( dz^2+\Lambda\cos^2(z)dy^2\right)\eqno(48')
$$
which can be recognized as a Kantowski-Sachs model. These models have a
$G_4$ symmetry which acts multiply-transitively. Although this solution
does not appear in the literature we assume that it must have been derived
by Kantowski in Ref.\cite{kantowski}, as mentioned in Ref.\cite{maccallum}.

In all these solutions the sign of the deceleration parameter is spatially
homogeneous, this is to say that it does not depend on the spatial coordinate
$z$. The results of this subsection can be easily generalized to the case of
trigonometric functions ($T(t)\propto\sin(at)$). In this case one can
speculate about the possibility of the periodic in time inflationary behaviour
- the universe in which the inflationary and non-inflationary periods
alternate each other in time.

\section{Qualitative Asymptotical Behaviour of the General Solution}
We will try to solve the general system of equations given by (11)-(15) with
the restrictions (18), (19), (20) and (21). As we have seen in Section III,
this leads to Eq.(23), from which the spatial part of the solution is Eq.(26).
The general solution is then given by Eq.(25). We have now three equations,
which are Eqs.(13)-(15). Eq.(13) becomes Eq.(23), as explained above, and the
new expressions for Eqs.(14) and (15), after introducing the general form of
the solution (25), are as follows:
\be
f'{Z\over Z'}={T\over\dot T}\left[{{\dot T\over T}(1+k^2-kmn)-\dot f
+(2m^2n-mk+n){\dot Q\over Q}}\right],
\ee
\begin{eqnarray}
-{Z''\over Z}-\left[{{1\over 2}n-{1\over 2}+m^2n^2+{k^2\over 4}- kmn}\right]
{Z'^2\over Z^2}&+&f'{Z'\over Z}\nonumber \\
={\ddot T\over T}+\left({k^2\over 4}-{1\over 2}\right){\dot T^2\over T^2}-
\dot f{\dot T\over T}+\left({{1\over 2} +m^2}\right){\dot Q^2\over Q^2}&-&km
{\dot T\over T}{\dot Q\over Q}.
\end{eqnarray}
We note that these equations are not completely separable, since the
derivatives of the function $f$ depend both on $t$ and $z$, which can be seen
from the following expression:
\be
f(t,z)={k^2\over 2}\log(TZ)-kmn\log(Z)-km\log(Q)+\log\left({{\ddot T
\over T}-{Z''\over Z}\over 2\Lambda}\right).
\ee
It can be shown, however, that for the three special cases under consideration,
$\dot f$ and $f'$ are functions of $t$ and $z$ alone, respectively.

Separating Eqs.(49) and (50) into a spatial and a time-dependent part and
introducing the function $\dot f$ into the time-dependent part of the Eq.(50),
we finally get
\be
\frac{\ddot T}{T}-\left(\frac{k^2}{4}+\frac{3}{2}\right)\frac{\dot T^2}{T^2}+
\left(\frac{1}{2}+m^2\right)\frac{\dot Q^2}{Q^2}-\left( 2m^2n+n\right)
\frac{\dot T}{T}\frac{\dot Q}{Q}=\sigma,
\ee
where $\sigma$ is a separation constant. Introducing now two new variables
\be
\alpha=\frac{\dot T}{T}\qquad\hbox{and}\qquad\beta=\frac{\dot Q}{Q},
\ee
the pertinent equations (23) and (52) can be written as:
\be
\left\{
\begin{array}{l}
\beta'+\alpha\beta=n\varepsilon a^2, \\*[3pt]
\alpha'-K\alpha^2+M\beta^2-2Mn\alpha\beta-\sigma=0,
\end{array}
\right.
\ee
where
\be
K\equiv\frac{k^2}{4}+\frac{1}{2}\qquad\hbox{and}\qquad M\equiv\frac{1}{2}+m^2.
\ee

Expressing $\alpha$ from the first equation and substituting it into the
second, we obtain a single equation for $\beta$. Introducing then the new
variable $y(\beta)=\beta'$, we get a non-linear differential equation of the
form
\be
y'y={\cal F}_0(\beta)+{\cal F}_1(\beta)y+{\cal F}_2(\beta)y^2,
\ee
where ${\cal F}_i(\beta)$ are certain non-linear functions of $\beta$.
This is a so-called Abel equation of the second type \cite{kamke}, whose
solution can not be inverted to obtain $\beta$ as a function of $y$. Therefore,
it is not possible to get the desired general solution for the system of
Eqs.(49), (50). However, we can try to make an analytical study of the system
(54) and get some indications of how the behaviour of the solutions is.

To achieve this, we have studied the phase space of the system (54), which is
equivalent to that of Eq.(56). The system has two equilibrium points which
are saddle points, therefore unstable. These two fixed points correspond to
inhomogeneous solutions satisfying $Q(t)=e^{\beta_0 t}$ and
$T(t)=e^{\alpha_0 t}$, with $\alpha_0$ and $\beta_0$ being the coordinates
of these points in the phase space. In Figures 1 and 2 we represent the
phase space of the system (54) for different values of the constants. The
trajectories of the solutions have a slightly different behaviour in both
cases, since in Figure 2 the two saddle points are closer to the $\beta= 0$
axis. In both figures, we have assumed that $\varepsilon =1$ in (54). In
the case that $\varepsilon =0$ and $\sigma =0$, which corresponds to
solutions linear in $z$, there is a special family of stable solutions,
those lying on the negative part of the $\alpha$-axis, that is, with
$\beta_0 =0$ and $\alpha_0\leq 0$. All these solutions evolve towards a
fixed point at the origin, of a static nature. This behaviour is shown in
Figure 3.

The same study can be done interchanging $t$ and $z$ coordinates, obtaining
the same system (54) as the final result, but this time $\alpha={Z'\over Z}$
and $\beta={P'\over P}$, and we have a system of equations depending
exclusively on the spatial coordinate.

\section{Conclusions}
We have studied a large class of $G_2$ inhomogeneous universes with a scalar
field with a potential of the Liouville type. After writing the Einstein
equations we have seen that there are two main classes of solutions. In the
first case, we have supposed that all the metric functions have a similar
spatial dependence. As a result of this assumption, we obtain that the
function $Z(z)$, which gives this spatial dependence, must have the form of
one of the three cases stated in Eq.(26). Another consequence is that the
sign of the deceleration parameter, which informs us about the existence of
inflation in these models depends only on spatial coordinate $z$, that is,
some parts of the universe will inflate while the other will not, at any time.
On the other hand, the second type of solutions represents somewhat an
opposite case: all the metric functions present a similar dependence in time,
and the function $T(t)$, which drives this dependence, is given by Eq.(37).
The sign of the deceleration parameter depends now on time only, so that the
whole of the universe inflate at the same time, but the periods of the
inflation and the duration depend on the initial conditions.

We have looked at the asymptotic behaviour of the solutions at late times
studying the system of equations (54). We have found no stable equilibrium
points for this system.

This work mainly was concerned with exact solutions. We have seen that by
introducing the large degree of inhomogeneity the cosmological model can have
a very complicated structure.

\section{Acknowledgements}
We are grateful to Prof. J.M.Aguirregabiria for assistance with the ODE
program \cite{ode}, which we have used to compute the asymptotical behaviour
of the solutions. P.L.'s work is supported by the Basque Government
fellowship B.F.I.92/090. This work is supported by the Spanish Ministry of
Education Grant (CICYT) No. PB93-0507.


\newpage
\thispagestyle{empty}
\begin{center}
{\bf FIGURE CAPTIONS}
\end{center}
\vskip 2cm
\noindent FIGURE 1: Phase space corresponding to the system of equations
(54) and the following values for the constants: $n=1$, $a=1$, $k=\sqrt{2}$,
$m=\sqrt{2}/2$ and $\sigma =1$. $\varepsilon$ is equal to $1$, so we are in
the hyperbolic case of section III.1.1.
\vskip 1cm
\noindent FIGURE 2: Phase space corresponding to the same system (54), but
for these values of the constants: $n=1/2$, $a=1$, $k=2$, $m=1$ and
$\sigma =0$. $\varepsilon$ is still $1$.
\vskip 1cm
\noindent FIGURE 3: Phase space corresponding to $\varepsilon =0$, that is,
to the linear case of section III.1.1. There is a collision of saddle
points at the origin. The values for the constants are those of Figure 1,
but with $\sigma =0$.

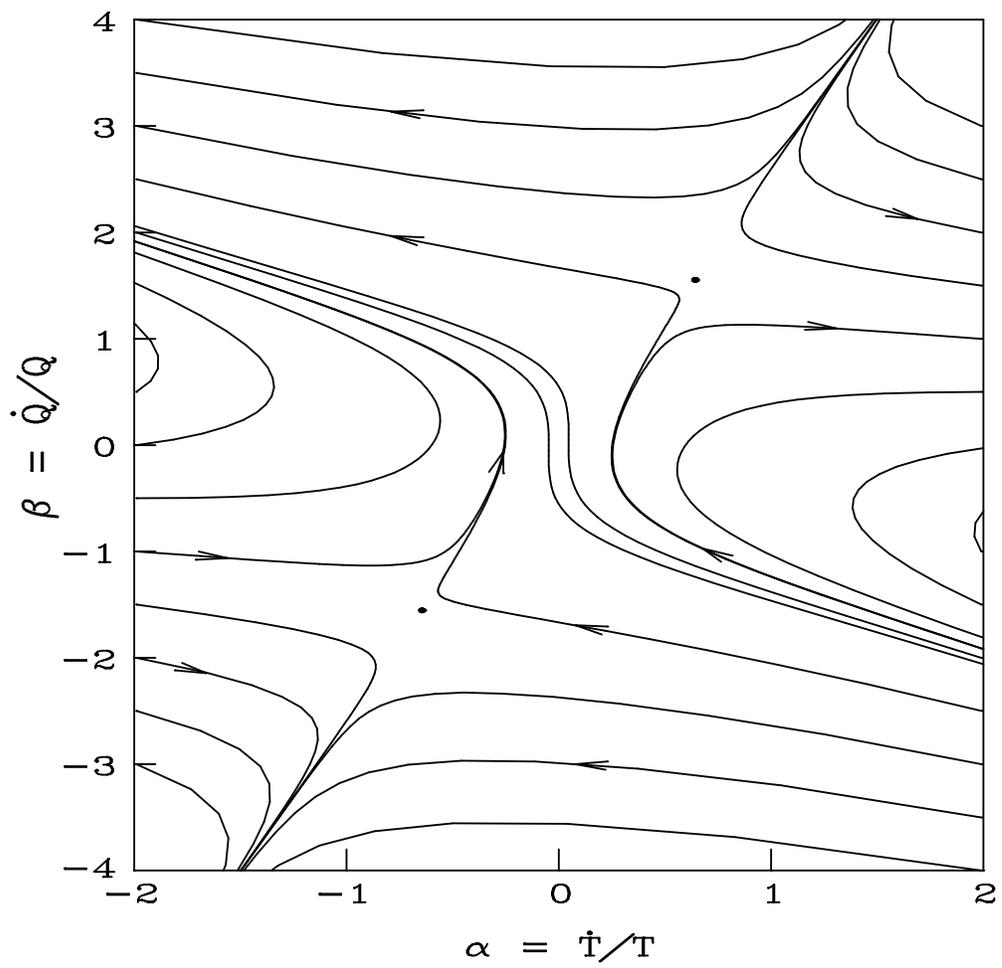
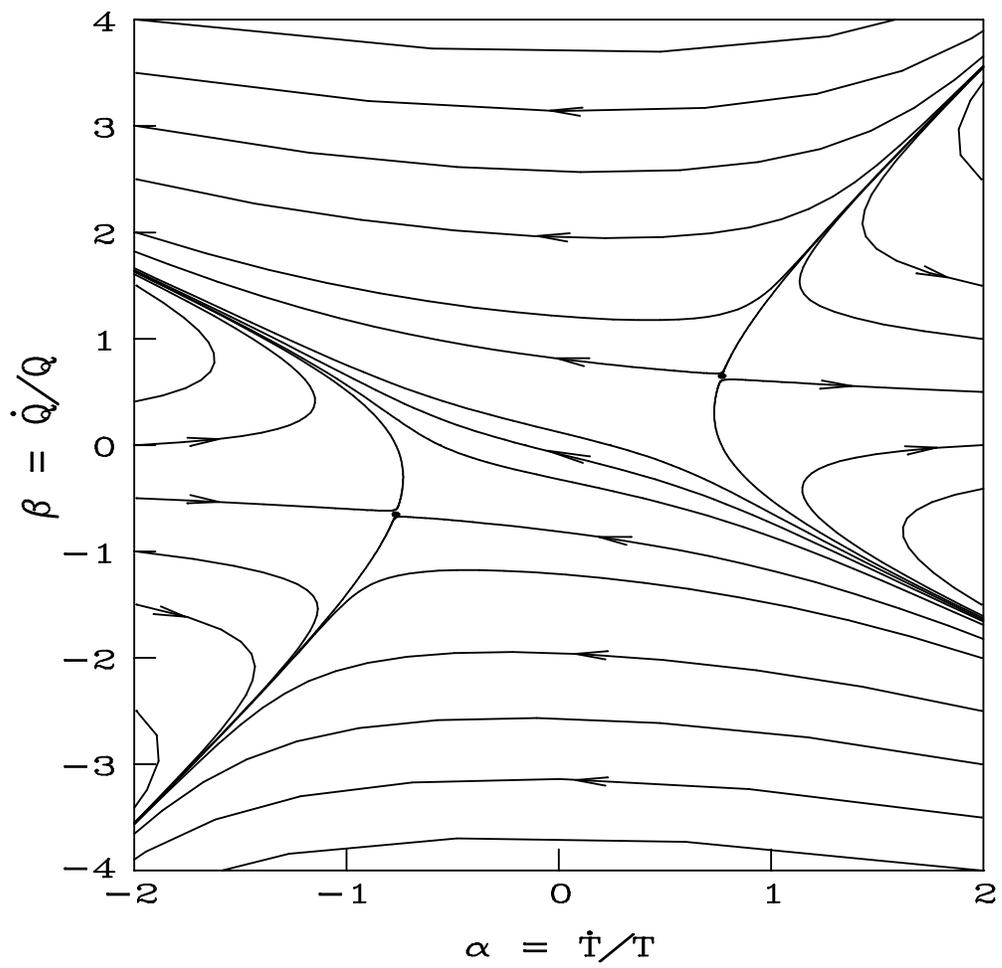
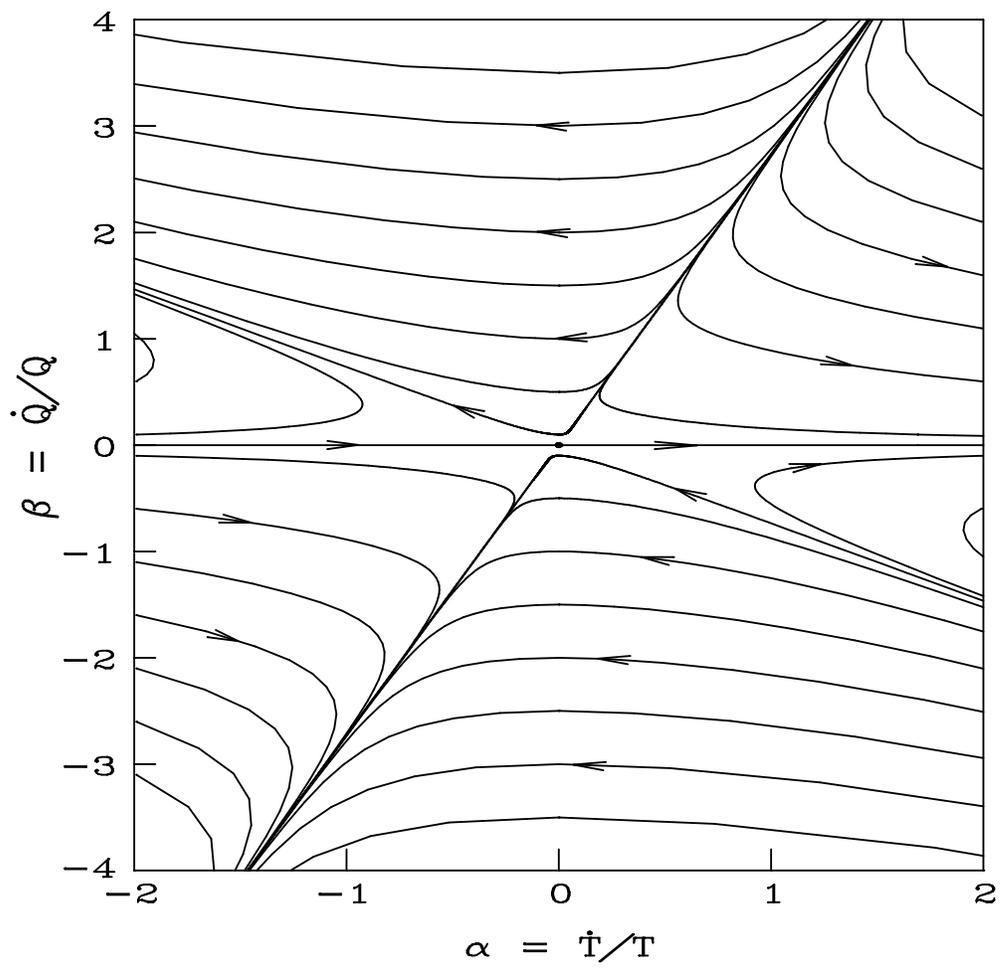
\end{document}